\documentclass[aps,prd,twocolumn,groupedaddress,showpacs]{revtex4}
\usepackage{float,graphicx}

\begin{document}


\title{Big Bang Nucleosynthesis with an Evolving Radion in the Brane World Scenario}



\author{B.~Li }
\email[Email address: ]{bli@phy.cuhk.edu.hk}
\affiliation{Department of Physics, The Chinese University of Hong
Kong, Hong Kong SAR, China}

\author{M.~-C.~Chu}
\email[Email address: ]{mcchu@phy.cuhk.edu.hk}
\affiliation{Department of Physics, The Chinese University of Hong
Kong, Hong Kong SAR, China}


\date{\today}

\begin{abstract}
We consider the big bang nucleosynthesis (BBN) in the Brane world
scenario, where all matter fields are confined on our 3-brane and
the radion of the Brane evolves cosmologically. In the Einstein
frame fundamental fermion masses vary and the results of standard
BBN (SBBN) are modified. We can thus use the observational
primordial element abundances to impose constraints on the
possible variations of the radion. The possibility of using the
evolving radion to resolve the discrepancies between the Wilkinson
Microwave Anisotropy Probe (WMAP) and SBBN values of the
baryon-to-photon ratio ($\eta$) is also discussed. The results and
constraints presented here are applicable to other models in which
fundamental fermion masses vary.
\end{abstract}

\pacs{26.35.+c, 98.80.Cq, 06.20.Jr, 11.10.Kk}

\maketitle


\section{INTRODUCTION}

The Brane world scenarios have been attracting much attention
during recent years, being motivated by the developments in
superstring theory and acting as alternatives to supersymmetric or
technicolor models in addressing the gauge hierarchy problem. In
such scenarios the matter fields are confined in our world, which
is believed to be a 3-dimensional brane embedded in a
$1+3+n$-dimensional bulk spacetime, while gravitation fills all
the dimensions. In Ref.~\cite{ADD1999} Arkani-Hamed, Dimopoulos
and Dvali (ADD) proposed a model in which the fundamental energy
scale can be as low as TeV while the weakness of gravity is due to
the large compactification radius of extra dimensions (millimeter
in case of two extra dimensions, for example). A different setup
was introduced by Randall and Sundrum (RS) \cite{RS1999} later, in
which our brane is one (the negative tension one) of the two
boundaries of a 5-dimensional anti-de Sitter $(\mathrm{AdS}_{5})$
spacetime and the gravitation is localized around the positive
tension brane while observers on our brane only measure a weak
tail of it. In some works, \emph{e.g.,} \cite{Dienes1998,
Chang2000, Appelquist2001}, the various matter fields are also
allowed to propagate along the extra dimensions and interesting
consequences are discovered such as a different unification scale
and the localization of zero-mode fermions onto our brane.

Brane model cosmology has also been investigated
extensively (see, \emph{e.g.,} \cite{Binetruy2000, Brax2003} for
an introduction). In particular, it is found \cite{Shiromizu2000} that
the gravitational equations on our 3-brane (in a 5-dimensional
bulk, for example) appear in the form
\begin{equation}
^{(4)}\mathrm{G}_{\mu\nu}=-\Lambda_{4}g_{\mu\nu}+8\pi
G_{\mathrm{N}}\tau_{\mu\nu}+\frac{64\pi^2}{M_{\mathrm{Pl}}^{*6}}\pi_{\mu\nu}-E_{\mu\nu},
\end{equation}
where $\tau_{\mu\nu}$ is the energy momentum tensor confined on
the brane and $M_{\mathrm{Pl}}^{*}$ the fundamental
(5-dimensional) Planck mass; $\Lambda_{4}$ is the effective
4-dimensional cosmological constant which is related to the bulk
cosmological constant $\Lambda$ and the brane tension $\lambda$ by
\begin{equation}
\Lambda_{4}=\frac{4\pi}{M_{\mathrm{Pl}}^{*3}}\left(\Lambda+\frac{4\pi}{3M_{\mathrm{Pl}}^{*3}}
\lambda^{2}\right).
\end{equation}
The 4-dimensional Newtonian constant $G_{\mathrm{N}}$ is defined
by
\begin{equation}
G_{\mathrm{N}}=\frac{4\pi\lambda}{3M_{\mathrm{Pl}}^{*6}}.
\end{equation}
The $\pi_{\mu\nu}$ term in Eq.~(1) is a quadratic function of
$\tau_{\mu\nu}$ and negligible when the energy density on the
brane is much smaller than the brane tension, \emph{i.e.},
$\tau_{\mu\nu}/\lambda\ll1$, and the last term is a part of the
high dimensional Weyl tensor carrying information of the
gravitation outside the brane, which acts in 4-dimensional theory
as a relativistic energy component $-$ the so-called dark
radiation. If the non-conventional field equations hold down to
low energies and the dark radiation term is nonzero, then the
predictions of the standard cosmological model, such as BBN and
CMB (Cosmic Microwave Background), might be modified dramatically.
Consequently the observations of primordial element abundances and
CMB power spectrum would place stringent constraints on the
parameters in Eq.~(1) \cite{Ichiki2002, Bratt2002}.

In Ref.~\cite{Bratt2002} the authors considered the effects of a
non-standard cosmic expansion rate on the outputs of BBN; yet
there is another source for modifying the SBBN yields in the Brane
World scenario -- namely the variation(s) of the fundamental
constant(s). This is because, although in most extra dimensional
models the moduli fields (for instance the radion in the brane
model) are believed to have been stabilized before the
commencement of the primordial nucleosynthesis, it is not
necessary to dismiss the possibility of cosmological evolution of
the moduli fields in order to avoid the long range forces and
deviations from general relativity (GR) \cite{Dent2003b}. It is
conceivable that there is a cosmological attractor mechanism as in
some scalar-tensor theories, driving the model towards the
conventional GR \cite{Brax2003a}; it is also probable that the
radion itself is a chameleon field \cite{Brax2004a}, acquiring a
large effective mass via its self-interaction together with the
interaction with matter and thus evading the constraints from
local gravitational measurements while evolving cosmologically
\cite{Brax2004b, Khoury2004}. Such a time evolution of the radion
field will lead to variations of fundamental constant(s).  BBN
turns out to impose the most stringent constraints on the possible
time evolution of the extra dimensions because it is more
sensitive than, say, CMB to the changes of the fundamental
constants. Therefore, it is our interest in the present work to
find out how the outputs of BBN are influenced by an evolving
radion field (See \cite{Campbell1995, Ichikawa2002, Li2005} for
the BBN constraints on other specific models and \cite{BIR1999,
Avelino2001, NL2002, Kneller2003, Flambaum2002} \emph{etc.~}for
constraints on the variations of individual fundamental constants
such as the fine structure constant and the strong coupling).

This article is arranged as following: in Sec.~II we discuss how
an evolving radion field changes the fundamental constant(s) and
thus leads to modified predictions of BBN. We will work in the
Einstein frame where $G_{\mathrm{N}}$ stays unaltered and show
that the only one varying fundamental constant in the model is the
Higgs vacuum expectation value (VEV) $v=\langle H\rangle$. The BBN
constraints on a changing Higgs VEV have been investigated
previously in \cite{Dixit1988, Scherrer1993} and more recently in
\cite{Yoo2003}, but ours differs from the previous works in
several ways: firstly, we consider some effects ignored in
previous works, such as the radiative and Coulomb corrections to
the neutron lifetime (and weak interaction rates), their radion
dependence and the radion dependence of the $p(n, D)\gamma$ cross
section; secondly, we have done a complete likelihood analysis
using the recent compilation on the nuclear rates and
uncertainties \cite{NACRE1999, Cyburt2000} and measurements of the
primordial abundances of $\mathrm{D}$, $^{4}\mathrm{He}$ and
$^{7}\mathrm{Li}$; thirdly, we show that the $^{7}\mathrm{Li}$
yields could be changed significantly with a small evolution of
the radion, tending to reduce the inconsistency between the SBBN and
WMAP-implied values of $\eta$, the baryon-to-photon ratio \cite{Spergel2003}.
Because of the accuracy of the WMAP result, we also use it to constrain the
variation of the radion. Our numerical results obtained from a
modified BBN code is presented in Sec.~III and then Sec~IV is
devoted to a discussion. Throughout this work we will assume three
species of massless neutrinos (or negligible neutrino masses) and
zero chemical potentials for neutrinos as in SBBN.

\section{THE INFLUENCES OF AN EVOLVING RADION ON BBN}

In this section we discuss the influences of an evolving radion on
the primordial nucleosynthesis. We first write down the low
energy action in the 4-dimensional effective theory and show that
only the Higgs VEV (and thus fundamental fermion masses) changes.
For simplicity we will concentrate on large flat extra dimensions;
the more realistic models with warped extra dimensions are
discussed in the literature, \emph{e.g.}, \cite{Brax2003a}. (In
\cite{Brax2003a} the authors considered a general class of two
brane models, including the RS model as a special case; the time
dependence of fermion masses in the Einstein frame and time
independence of gauge couplings can also be found there.) Then we
briefly discuss the implications on BBN. The derivations in this
section closely follow our recent work \cite{Li2005}.

\subsection{The Low Energy Effective Action}

Let us start from a general $4+n$-dimensional model, with $n$
being the number of the large extra dimensions. The full line
element is given as:
\begin{equation}
d\hat{s}^{2} =
G_{AB}dX^{A}dX^{B}=g_{\mu\nu}dx^{\mu}dx^{\nu}+h_{ab}dy^{a}dy^{b},
\end{equation}
where $\mu$, $\nu$ = 0, 1, 2, 3 label the four ordinary
dimensions, $a$, $b$ = $4, \cdot\cdot\cdot, 3+n$ denote extra
dimensions and $A, B = 0, 1, 2, \cdot\cdot\cdot, 3+n$ describe the
whole spacetime. (We shall not consider cross terms such as
$G_{a\mu}$ in Eq.~(4)). The extra dimensions are assumed to
compactify on an orbifold, and their coordinates $y_{a}$ take
values in the range [0, 1]. The quantities $h_{ab}$ have
dimensions of [Length]$^{2}$ since $y_{a}$ are dimensionless in
our choice.

Then the effective 4-dimensional action in the gravitational
sector can be obtained by dimensionally reducing Eq.~(4) as:
\begin{widetext}
\begin{eqnarray}
S_{\mathrm{Gravity}} &=& \frac{1}{\kappa_{4+n}^{2}}\int
d^{4+n}X\sqrt{|G|}R_{4+n}[G]\nonumber\\ &=&
\frac{1}{\kappa_{4}^{2}}\int d^{4}xd^{n}y \sqrt{|g|}
\frac{\sqrt{|h|}}{V_{0}}\left[R_{4}\left[g\right]-
\frac{1}{4}\partial_{\mu}h^{ab}\partial^{\mu}h_{ab}
-\frac{1}{4}h^{ab}\partial_{\mu}h_{ab}\cdot
h^{cd}\partial^{\mu}h_{cd}\right], \label{eq:wideeq}
\end{eqnarray}
\end{widetext}
in which $|g|$, $|h|$ and $|G|$ are respectively the determinants
of the metrics of the ordinary dimensions, the extra dimensions
and the whole spacetime. $R_{4}[g]$ and $R_{4+n}[G]$ are the Ricci
scalars of the ordinary 4 and the total $4+n$ dimensional
spacetimes. $\kappa_{4}, \kappa_{4+n}$ are related to the 4 and
$4+n$ dimensional Planck masses through $\kappa_{4}^{2} = 2
M_{\text{Pl},4}^{2}$ and $\kappa_{4+n}^{2} = 2
M_{\text{Pl},4+n}^{2+n}$, while they themselves are connected by a
volume suppression $\kappa_{4+n}^{2} = \kappa_{4}^{2}\cdot V$, $V$
being a measure of the extra space volume whose present-day value
is denoted by $V_{0}$ in Eq.~(5) (Note that because of the
specified choice of $V_{0}$ and because the higher dimensional
quantity $\kappa_{4+n}$ is treated as a constant, the $\kappa_{4}$
above also takes its currently measured value and is a constant
rather than a variable).

The effective Ricci curvature term is not canonical in Eq.~(5); to
make it so, let us take the conformal transformation
\begin{equation}
g_{\mu\nu} \rightarrow e^{2\vartheta} g_{\mu\nu}
\end{equation}
and choose the field $\vartheta$ to satisfy
\begin{equation}
\frac{\sqrt{|h|}}{V_{0}}e^{2\vartheta} = 1.
\end{equation}
Then we obtain the effective 4-dimensional gravitational action in
the Einstein frame:
\begin{widetext}
\begin{eqnarray}
S_{\mathrm{Gravity}} = \frac{1}{\kappa_{4}^{2}}\int
d^{4}x\sqrt{|g|} \left[R_{4}-
\frac{1}{4}\partial_{\mu}h^{ab}\partial^{\mu}h_{ab} +
\frac{1}{8}h^{ab}\partial_{\mu}h_{ab}\cdot
h^{cd}\partial^{\mu}h_{cd}\right]. \label{eq:wideeq}
\end{eqnarray}
\end{widetext}

We shall make a further assumption that the extra dimension(s) are
homogeneous and isotropic, \emph{i.e}., the metric of the extra
space takes the following form:
\begin{equation}
h_{ab} = \mathrm{diag} (-b^{2}, -b^{2}, \cdot\cdot\cdot, -b^{2}),
\end{equation}
and then the action Eq.~(8) could be rewritten as
\begin{equation}
S_{\mathrm{Gravity}} = \int
d^{4}x\sqrt{|g|}\left[\frac{1}{\kappa_{4}^{2}}R_{4}+\frac{1}{2}
g^{\mu\nu}\partial_{\mu}\sigma\partial_{\nu}\sigma\right]
\end{equation}
by defining a new scalar field, the radion $\sigma$:
\begin{equation}
\sigma \equiv
\frac{1}{\kappa_{4}}\sqrt{\frac{n+2}{n}}\mathrm{log}\frac{b^{n}}{V_{0}}.
\end{equation}

Next we turn to the matter sector of the effective action, firstly
for the scalar fields. The action of a brane scalar field $\phi$
is given by:
\begin{equation}
S_{\phi} = \int d^{4}x\sqrt{|g|}
\left[\frac{1}{2}g^{\mu\nu}\partial_{\mu}\phi
\partial_{\nu}\phi-\hat{U}(\phi)\right],
\end{equation}
because the $n$ large extra dimensions are inscient to the field
and have been integrated out. Then the same conformal
transformation given in Eqs.~(6) and (7) transforms Eq.~(12) into
the following form:
\begin{widetext}
\begin{equation}
S_{\phi} = \int d^{4}x\sqrt{|g|}
\left\{\frac{1}{2}\text{exp}\left[-\kappa\sqrt{\frac{n}{n+2}}
\sigma\right]g^{\mu\nu}\partial_{\mu}\phi
\partial_{\nu}\phi-\text{exp}\left[-2\kappa
\sqrt{\frac{n}{n+2}}\sigma\right]\hat{U}(\phi)\right\}.
\label{eq:wideeq}
\end{equation}
\end{widetext}
Note that from now on we will use $\kappa$ instead of $\kappa_{4}$
for simplicity. The kinetic term of the scalar field in Eq.~(13)
could be made canonical by redefining a new scalar field as
\begin{equation}
\varphi\equiv\text{exp}\left[-\frac{1}{2}\kappa\sqrt{\frac{n}{n+2}}
\sigma\right]\phi,
\end{equation}
and the action becomes (up to the higher order derivative term):
\begin{equation}
S_{\varphi} = \int d^{4}x\sqrt{|g|}
\left[\frac{1}{2}g^{\mu\nu}\partial_{\mu}\varphi
\partial_{\nu}\varphi-U(\varphi)\right],
\end{equation}
where the potential $U(\varphi)$ is given by
\begin{equation}
U(\varphi)=\frac{1}{2}\text{exp}\left[-\kappa\sqrt{\frac{n}{n+2}}
\sigma\right]m_{\phi}^{2}\varphi^{2}+\lambda\varphi^{4}
\end{equation}
provided that the original potential $\hat{U}(\phi)$ takes the
following form:
\begin{equation}
\hat{U}(\phi)=\frac{1}{2}m_{\phi}^{2}\phi^{2}+\lambda\phi^{4}.
\end{equation}

The same technique could be applied to gauge fields, whose action
before conformal transformation is given as
\begin{equation}
S_{\text{Gauge}} = -\int d^{4}x\sqrt{|g|}
\frac{1}{4\tilde{g}^{2}}\hat{F}^{r\mu\nu}\hat{F}_{\mu\nu}^{r},
\end{equation}
where $\tilde{g}$ is the gauge coupling constant (a tilde is used
to distinguish it from $g$, the determinant of the
metric) and $\hat{F}_{\mu\nu}^{r}$ are corresponding gauge field
strengths. With the conformal transformation Eq.~(6) the above
action is changed into
\begin{equation}
S_{\text{Gauge}} = -\int d^{4}x\sqrt{|g|} \frac{1}{4\tilde{g}^{2}}
F^{r\mu\nu}F_{\mu\nu}^{r},
\end{equation}
where $\tilde{g}$ is unchanged because
$F_{\mu\nu}=\hat{F}_{\mu\nu}$ have zero conformal weights and are
unaltered under conformal transformations.

We can also obtain the effective action for the Dirac fermion
field in a similar way. Starting from the brane action:
\begin{equation}
S_{\Psi} = \int d^{4}x \sqrt{|g|}
\left\{i\bar{\Psi}\gamma^{\mu}D_{\mu}\Psi-\hat{m}\bar{\Psi}\Psi\right\}
\end{equation}
and using the conformal transformation Eq.~(6) we get
\begin{widetext}
\begin{eqnarray}
S_{\Psi} = \int d^{4}x \sqrt{|g|}
\left\{\text{exp}\left[-\frac{3\kappa}{2}\sqrt{\frac{n}{n+2}}
\sigma\right]i\bar{\Psi}\gamma^{\mu}D_{\mu}\Psi
-\text{exp}\left[-2\kappa\sqrt{\frac{n}{n+2}}
\sigma\right]\hat{m}\bar{\Psi}\Psi\right\}. \label{eq:wideeq}
\end{eqnarray}
\end{widetext}
To make the kinetic part of the fermion action Eq.~(21) canonical,
we rescale the field as
\begin{equation}
\psi \equiv
\mathrm{exp}\left[-\frac{3\kappa}{4}\sqrt{\frac{n}{n+2}}\sigma\right]\Psi,
\end{equation}
and then using the conformality of the coupling of massless Weyl
fermions, we rewrite Eq.~(21) as:
\begin{equation}
S_{\psi} = \int d^{4}x \sqrt{|g|}
\left\{i\bar{\psi}\gamma^{\mu}D_{\mu}\psi
-m(\sigma)\bar{\psi}\psi\right\},
\end{equation}
in which
\begin{equation}
m(\sigma)\equiv\text{exp}\left[-\frac{\kappa}{2}\sqrt{\frac{n}{n+2}}
\sigma\right]\hat{m}.
\end{equation}
Note that our results above are equal to those of
\cite{Mazumdar2004} when there are no universal extra dimensions
(\emph{i.e.}, $n=0$ in their model).

It is apparent from Eq.~(24) that if the radion $\sigma$ evolves,
then the fermion masses also vary. As in the standard model, we
assume that the fundamental fermion masses are generated by the
Higgs mechanism and that the Higgs potential takes the form as
Eq.~(17). We find from Eq.~(16) that the Higgs VEV ($v$), which is
obtained by simply minimizing the Higgs potential, has the same
radion dependence as the fermion masses (Eq.~(24)) so that the
Yukawa couplings are independent of radion. Recalling also that
the gauge couplings would not depend on radion (Eq.~(19)), which
is because of, as pointed above, the conformal invariance of the
gauge kinetic term and this is true in general scalar tensor
gravity theories \cite{Campbell1995}. So here we indeed encounter
a model in which the only varying fundamental constant is $v$. On
the other hand, if the gauge fields exist in the full dimensions
as suggested by the universal extra dimensional scenario
\cite{Appelquist2001}, then the above conclusion does not hold
anymore and the gauge couplings will be dependent on the radion
field \cite{Li2005}.

\subsection{BBN with a Varying Higgs VEV}

The primordial nucleosynthesis with a varying Higgs VEV has been
discussed previously in \cite{Dixit1988, Scherrer1993, Yoo2003},
and here we shall only briefly identify the places where effects
of Higgs VEV enter. We also omit the introduction to the BBN
theory and observation and refer these to existing literatures,
\emph{e.g.}, \cite{Kolb1990, SKM1993, Sarkar1996, Olive2000a,
Olive2000b, Tytler2000, Serpico2004}.

A different $v$ from its standard value will modify BBN mainly
in two aspects, the weak interactions and the nuclear reactions.
Consider firstly the weak interactions
\begin{eqnarray*}
\nu_{e} + n &\leftrightarrow& p + e^{-}, \nonumber\\
e^{+} + n &\leftrightarrow& p + \bar{\nu_{e}}, \nonumber\\
n &\leftrightarrow& p + e^{-} + \bar{\nu_{e}},
\end{eqnarray*}
which interconvert the neutrons and protons. Roughly speaking,
they are important for BBN because they determine the neutron
density at the beginning of BBN, and, because nearly all the
neutrons are incorporated into $^{4}\text{He}$ at last, they are
crucial for the final $^{4}\text{He}$ output (for more detailed
discussions see \emph{e.g.}, \cite{Kneller2003}). The rates of
these $n \leftrightarrow p$ interactions could be well described
as
\begin{widetext}
\begin{eqnarray}
\Gamma(n\rightarrow p) = &A& \int_{1}^{\infty} d\epsilon
\frac{\epsilon(\epsilon-q)^{2}(\epsilon^{2}-1)^{1/2}}{\left
[1+\mathrm{exp}(-\epsilon z_{e})\right ]\left
\{1+\mathrm{exp}\left [(\epsilon -q)z_{\nu}\right ]\right
\}}\nonumber\\ &+& A \int_{1}^{\infty} d\epsilon
\frac{\epsilon(\epsilon+q)^{2}(\epsilon^{2}-1)^{1/2}}{\left
[1+\mathrm{exp}(\epsilon z_{e})\right ]\left \{1+\mathrm{exp}\left
[-(\epsilon +q)z_{\nu}\right ]\right \}}, \label{eq:wideeq}
\end{eqnarray}
\begin{eqnarray}
\Gamma(p\rightarrow n) = &A& \int_{1}^{\infty} d\epsilon
\frac{\epsilon(\epsilon-q)^{2}(\epsilon^{2}-1)^{1/2}}{\left
[1+\mathrm{exp}(\epsilon z_{e})\right ]\left \{1+\mathrm{exp}\left
[(q-\epsilon)z_{\nu}\right ]\right \}}\nonumber\\
&+& A\int_{1}^{\infty}
d\epsilon\frac{\epsilon(\epsilon+q)^{2}(\epsilon^{2}-1)^{1/2}}{\left
[1+\mathrm{exp}(-\epsilon z_{e})\right ]\left
\{1+\mathrm{exp}\left [(\epsilon +q)z_{\nu}\right ]\right \}},
\label{eq:wideeq}
\end{eqnarray}
\end{widetext}
using the Born approximation, where we have defined dimensionless
quantities $q=m_{np}/m_{e}$, $\epsilon=E_{e}/m_{e}$ and
$z_{\nu}=m_{e}/T_{\nu}$ and $z_{e}=m_{e}/T$ with $m_{e}$ being the
electron mass, $m_{np}$ the neutron-proton mass difference and
$T_{\nu}$ ($T$) the temperature of the neutrinos (the
electromagnetic plasma). Here $A$ is a normalization factor
determined by the requirement that at zero temperature $\Gamma(n
\rightarrow p + e^{-} + \bar{\nu_{e}})=\tau_{n}^{-1}$ with
$\tau_{n}$ the neutron lifetime, which means that:
\begin{equation}
A = \tau_{n}^{-1} \lambda(q)^{-1} \propto
G_{\text{F}}^{2}m_{e}^{5} ,
\end{equation}
where
\begin{equation}
\lambda(q) = \int_{1}^{q} d\epsilon\  \epsilon
(\epsilon-q)^{2}(\epsilon^{2}-1)^{1/2}.
\end{equation}

>From Eqs.~(25)-(28) we can see that $v$ determines the weak rates
through $G_{\text{F}}$, $m_{e}$ and $m_{np}$ respectively. The
dependences of $G_{\text{F}}$ and $m_{e}$ on $v$ can be simply
parameterized as
\begin{eqnarray}
G_{\text{F}} &\propto& \frac{1}{v^{2}},\\
m_{e} &\propto& v.
\end{eqnarray}
The neutron-proton mass difference $m_{np}$ is a sum of the
electromagnetic contribution ($\sim -0.76\ \text{MeV}$) and the
$u-d$ quark mass difference ($\sim 2.053\ \text{MeV}$)
\cite{Ichikawa2002}, for which the former is unchanged in the
present model because the (electromagnetic and strong) gauge
couplings are radion independent while the latter is proportional
to the Higgs VEV. So $m_{np}$ could be described as
\begin{equation}
m_{np}=2.053\rho\ \text{MeV}-0.76\ \text{MeV},
\end{equation}
in which
\begin{equation}
\rho \equiv \frac{v_{\text{BBN}}}{v_{\text{NOW}}}
\end{equation}
is the ratio between the Higgs VEVs at the BBN era and at present.
$m_{np}$ is important also because it determines the equilibrium
neutron-to-proton ratio through \cite{Kolb1990}
\begin{equation}
\frac{n_{n}}{n_{p}}=\text{exp}\left[-\frac{m_{np}}{T}\right].
\end{equation}

Although the Born approximation in Eqs.~(25) and (26) captures the
essential features, various corrections are needed for more
accurate estimations (see \emph{e.g.}, \cite{Dicus1982, Lopez1999,
Esposito1999, Serpico2004}). For example, the accuracy of the
theoretical value of $\tau_{n}$ from Born approximation is $\sim
7\%$ while including the zero-temperature and Coulomb corrections
reduces it to be $\sim 0.1\%$ \cite{Serpico2004}. These two
corrections are the most important ones and the radiative
correction itself depends on $m_{np}$ (and thus on $\rho$); in our
calculation we numerically integrate Eqs.~(25) and (26) and
include these corrections explicitly. We have neglected other
corrections for simplicity and as the accuracy is adequate for our
purpose here (see \cite{Serpico2004, Dicus1982, Lopez1999,
Esposito1999} for discussions on this issue).

Next we turn to the nuclear reaction sector. The change of $v$
will modify the pion mass $m_{\pi}$, which is related to the
light quark mass $m_{q}$ \cite{GMOR}:
\begin{equation}
m_{\pi} \propto m_{q}^{\frac{1}{2}} \propto \rho^{\frac{1}{2}}.
\end{equation}
According to recent
investigations of the dependence of the nuclear potential on the
pion mass \cite{Epelbaum2003, Beane2003}, the deuteron binding
energy $B_{d}$ shows a strong decrease when $m_{\pi}$ increases;
although the relationships between $B_{d}$ and $m_{\pi}$ as
obtained in \cite{Epelbaum2003, Beane2003} have large
uncertainties, they could be well approximated by
\begin{equation}
B_{d}=B_{d,0}\left[(r+1)-r\frac{m_{\pi}}{m_{\pi,0}}\right]
\end{equation}
within the small range of $m_{\pi}$ (or $\rho$) we are working
with \cite{Yoo2003, Muller2004}, in which $B_{d,0}$ and
$m_{\pi,0}$ denote their present-day values and the central value
of $r$ ranges from 6 \cite{Epelbaum2003} to 10 \cite{Beane2003}.
In the following we shall work with $r=10$ and $6$ in parallel and
show that the resulting qualitative features are the same.

\begin{figure}[]
\includegraphics[scale=1]{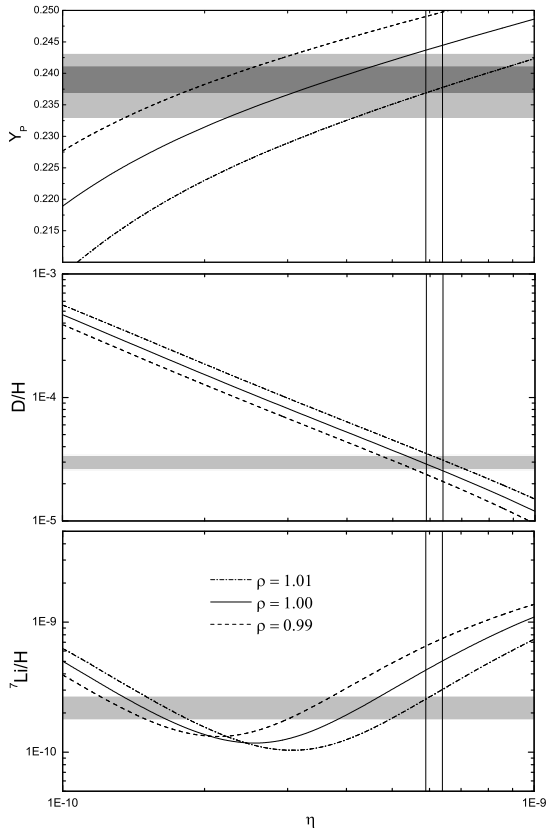}
\caption{The primordial abundances of $^{4}\mathrm{He}$, D and
$^{7}\mathrm{Li}$ as a function of the baryon-to-photon ratio
$\eta$, for various values of $\rho$ as indicated by the legend in
the lower panel and $r=10$. Also shown are the observational
$1\sigma$ ranges of the abundances of $^4\text{He}$ from
\cite{Olive2000b} (grey region) and \cite{Luridiana2003} (dark
grey region), $\text{D}$ from \cite {Kirkman2003}(grey region),
$^7\text{Li}$ from \cite{Bonifacio2002} (grey region), as well as
the baryon-to-photon ratio implied by WMAP \cite{Spergel2003},
$\eta_{\text{WMAP}}=(6.14\pm0.25)\times10^{-10}$ (the vertical
lines).}
\end{figure}

\begin{figure}[]
\includegraphics[scale=1]{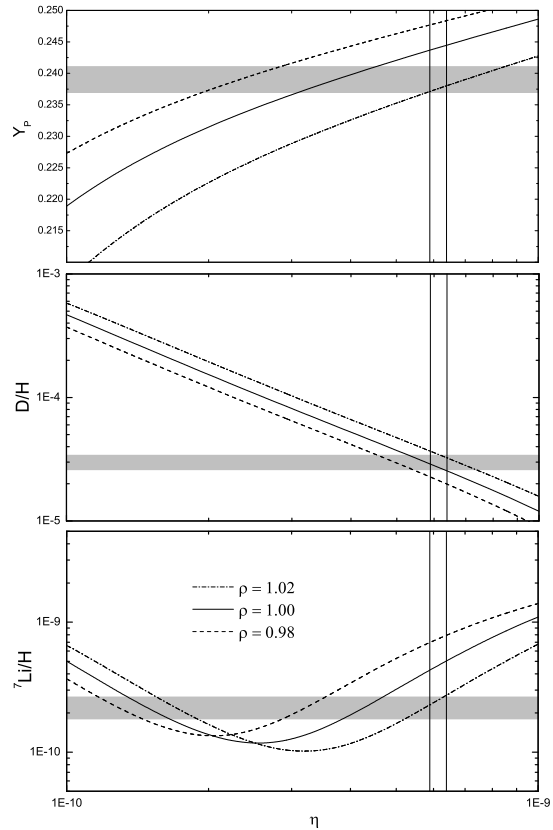}
\caption{Same as Fig.~1, but here $r=6$ and the values of $\rho$
are different.}
\end{figure}

The deuteron binding energy plays an important role in BBN because
it determines whether a significant amount of $\text{D}$ could be
produced (via the reaction $p(n, \text{D})\gamma$), which then leads
to the synthesis of heavier elements. Because of the huge number
of photons (recall that $\eta \sim 10^{-10}$!), the inverse
process $\gamma(\text{D}, n)p$ could be active down to low
energies ($\sim 0.08\ \text{MeV}$) and sensitive to the variations
in $B_{d}$ \cite{Kneller2003, Dmitriev2004}. Besides, the forward
cross section of $p(n, \text{D})\gamma$ will also depend on
$B_{d}$ \cite{Dmitriev2004}; to estimate this dependence, we adopt
the cross section formula calculated using the effective field
theory without pion \cite{Chen1999, Rupak2000}. The authors of
\cite{Chen1999, Rupak2000} gave an expression of the $p(n,
\text{D})\gamma$ cross section in terms of several parameters in
which the deuteron binding energy $B_{d}$ has by far the largest
leverage. In our modified BBN code we average the cross section
given in \cite{Chen1999} over the thermal distribution of the
particles numerically and obtain the reaction rate as a function
of the temperature \cite{Fowler1967}; we have checked that the
results obtained in this way are essentially identical to those
obtained using the fitted $p(n, \text{D})\gamma$ rates of
\cite{SKM1993, Cyburt2004}.

For the other reactions which are of
importance to the $\text{D}$ yields, such as $\text{D}(\text{D},
n){}^3\text{He}$, $\text{D}(\text{D},p){}^3\text{H}$, $\text{D}
(p,\gamma){}^3\text{He}$, $\text{D}({}^3\text{H},n){}^4\text{He}$
and $\text{D}({}^3\text{He},p){}^4\text{He}$, their cross sections
may also be dependent on $m_{\pi}$ or $B_{d}$ but we have no similar
effective field theory calculations on them; so we ignore them in
the present article. If we simply take these cross
sections to be related to the size of the deuteron radius,
\emph{i.e.~}$\sigma\propto1/B_{d}$, as in Ref.~\cite{Kneller2003},
then in the interested range of $\eta$ and $\rho$ (See FIG.~5 below)
we find modifications to the final outputs of $\text{D}$,
$^4\text{He}$ and $^7\text{Li}$ by $<\sim6\%$, $<\sim7\times10^{-4}$
and $<\sim3\%$ respectively, all well lying within the corresponding $1\sigma$
observational uncertainties ($\sim16\%$ for $\text{D}$, $\sim0.84\%$ for
$^4\text{He}$ and $\sim21\%$ for $^7\text{Li}$, see Eqs.~(36)-(38)
below). In the present work we do not adopt the $1/B_{d}$ parametrization of
these cross sections due to the lack of more explicit expressions;
rather we treat the above estimations as possible errors introduced by
neglecting their $m_{\pi}$ or $B_{d}$ dependences and emphasize
that further related theoretical calculations are needed to reduce these
errors.

Note that the variation of the Higgs VEV probably will also cause
modifications in the binding energies of other nuclei and thus
change the corresponding reverse reaction rates. However there are
again no explicit calculations of these modifications. Fortunately
these effects are small because the abundances of these heavier
nuclei fall far below their nuclear statistical equilibrium (NSE)
values long before the beginning of BBN and the reverse reactions
are effectively switched off. For example, according to our
constraints on $\rho$, \emph{i.e.,~}$\rho\sim1.01$ (see below),
the deuteron binding energy $B_{d}$ will be decreased by
$\sim5\%$. Let us suppose that the other binding energies have
variations of similar magnitude, since different nuclei may appear in the
left-hand and right-hand sides of a reaction
(\emph{e.g.,~}$^3\text{He}$ and $^3\text{H}$ in $^3\text{He}(n, p)
^3\text{H}$) and we do not know whether the variations of their
binding energies correlate or anti-correlate.  We thus arrived at a
conservative estimation of $\sim10\%$ for the Q-values of these
reactions. If the Q-values of all the reactions relevant to BBN
are changed by $10\%$, then the abundances of $\text{D}$,
$^4\text{He}$ and $^7\text{Li}$ would be varied by no more than
$\sim0.1\%$, $\sim5\times10^{-5}$ and $\sim0.5\%$ respectively, in
the range of $\eta$ of interest \cite{Comment2}.

There is also a modification to the cosmic expansion rate in the
present model. As the Newtonian constant $G_{\text{N}}$ is
unchanged in the Einstein frame, this modification originates from
the variation of energy density compared with SBBN --- due to the
modified electron and positron masses as described in
\cite{Yoo2003, Li2005}. We include this effect in the calculation
although it is small compared with others.

\begin{figure}[]
\includegraphics[scale=0.915]{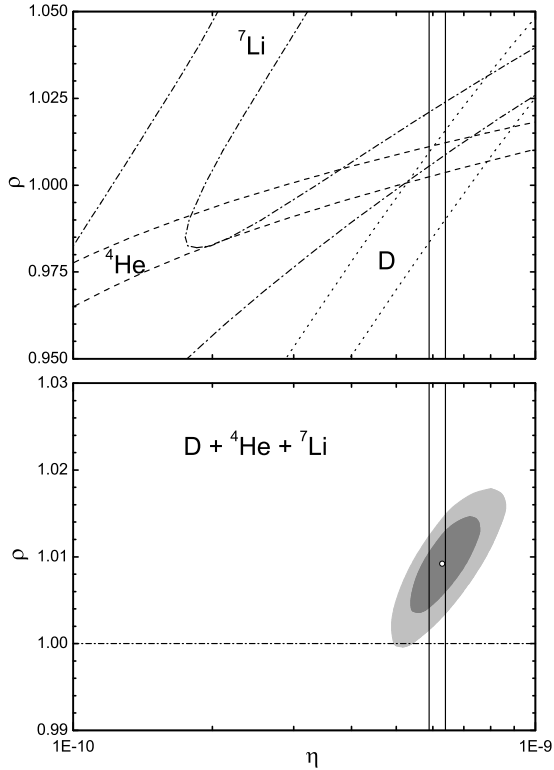}
\caption{Upper panel: individual contours of $\text{D}$ (dotted),
$^{4}\text{He}$ (dashed) and $^{7}\text{Li}$ (dot-dashed) at 68\%
confidence level. Lower panel: joint contours of
$\text{D}$+$^{4}\text{He}$+$^{7}\text{Li}$ at 68\% (dark grey
region) and 95\% (grey region) confidence levels; the best-fitting
parameters $(\eta, \rho)\simeq(6.28\times10^{-10}, 1.009)$ are
denoted by the white circle and the case of SBBN by the horizontal
dash-dotted line. In both panels the range of $\eta_{\text{WMAP}}$
is represented by vertical solid lines. Here $r$ is taken to be
10.}
\end{figure}

\begin{figure}[]
\includegraphics[scale=0.915]{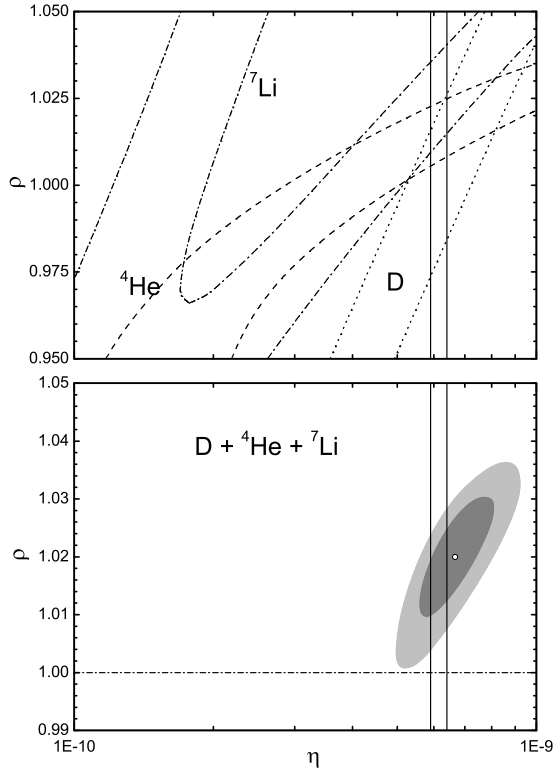}
\caption{Same as Fig.~3, but for $r=6$. The best fitting point
here is $(\eta, \rho)\simeq(6.67\times10^{-10}, 1.020)$.}
\end{figure}

\section{NUMERICAL RESULTS}

We have incorporated all the effects discussed in the previous
section into the standard BBN code by Kawano \cite{Kawano1992} and
used it to obtain the numerical results presented in this section.
The nuclear reaction rates are taken from the NACRE compilation
\cite{NACRE1999} (for the rates not given in \cite{NACRE1999} we
adopt those of \cite{SKM1993}) and their uncertainties from the
work of Cyburt, Fields and Olive \cite{Cyburt2000}. For the
present neutron lifetime we use the value suggested by the
Particle Data Group \cite{PDG2004}, $\tau_{n}^{\text{ex}}=885.7
\pm 0.8\ \text{s}$.

In Fig.~1 we plot the abundances of the light nuclei
$^{4}\text{He}$, D and $^{7}\text{Li}$ with a modified Higgs VEV
($v$), which is characterized by $\rho$, in the case of $r=10$. It
is apparent that if $v$ is larger at the BBN era $(\rho > 1)$,
then the deuterium output increases while the $^{4}\mathrm{He}$
and $^{7}\mathrm{Li}$ (for $^{7}\mathrm{Li}$ we only consider the
larger-$\eta$-case implied by the WMAP result
\cite{Spergel2003}) yields decrease compared with SBBN. The
behavior of the $^{4}\mathrm{He}$ output is a consequence of
several effects: firstly, a $\rho$-value larger than 1 leads to a
larger $m_{np}$ (from Eq.~(31)) and a smaller neutron density at
the beginning of the nucleosynthesis (from Eq.~(33)), thus finally
to a smaller $^{4}\mathrm{He}$ output; secondly, the weak
interaction rates will be smaller than those in SBBN, leading to
increased final $^{4}\mathrm{He}$ abundance.  Thirdly, the
deuteron binding energy $B_{d}$ becomes smaller (from Eq.~(35))
than in SBBN so that the nucleosynthesis will commence at a later
time and become less efficient, producing less $^{4}\mathrm{He}$
and $^{7}\mathrm{Li}$ while leaving more D unprocessed (there is
an extra decrease in the forward rate of $p(n, \text{D})\gamma$,
whose influence is small compared with the effect of commencing BBN later).
The sum of these effects is a smaller
$^{4}\text{He}$ abundance (see Fig.~1 of \cite{Yoo2003} for a more
explicit comparison among these effects). For comparison we also plot
in Fig.~2 the case of $r=6$, where the effect of $B_{d}$ becomes
weaker than $r=10$ but the essential features are the same.

\begin{figure}[]
\includegraphics[scale=0.82]{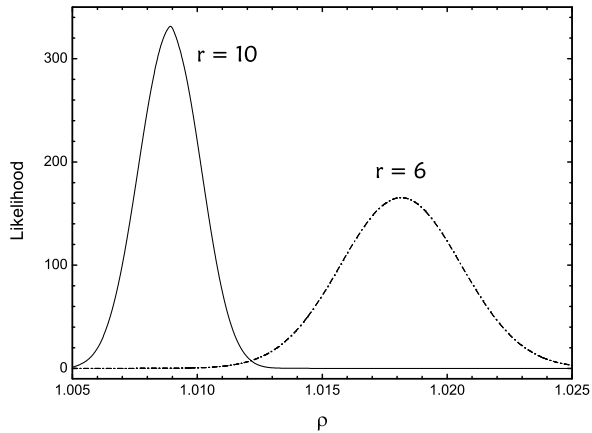}
\caption{The likelihood functions as functions of $\rho$ alone
with WMAP constraint on $\eta$ imposed, for $r=10$ (the solid
curve) and $r=6$ (the dot-dashed curve). Both curves are
normalized such that the areas under them are unity.}
\end{figure}

The results in Figs.~1 and 2 suggest that the discrepancy between
the SBBN theory and WMAP observations is reduced if $\rho$ is
greater than 1 \cite{Comment1}. It is well known that adopting the
WMAP-implied value of $\eta$ \cite{Spergel2003}, the SBBN can
reproduce the observed D abundance, but more $^{4}\mathrm{He}$ and
$^{7}\mathrm{Li}$  than their observational abundances. One
possibility is that this discrepancy is due to some systematic
errors in the $^{4}\mathrm{He}$ and $^{7}\mathrm{Li}$
observations. However, it may also originate from some new
physics, such as a varying fine structure constant \cite{NL2002},
an altered deuteron binding energy \cite{Dmitriev2004}, more than
three relativistic (effective) neutrino species
\cite{Barger2003a}, a lepton asymmetry \cite{Barger2003b}, or some
combination of them \cite{Ichikawa2004}, \emph{etc}. Our results
show that we might add another candidate, a varying Higgs VEV, to
this list. To see this point more quantitatively, and to derive a
constraint on the possible evolution of the radion, next we will
give the likelihood analysis of this model. For this purpose, we
choose to use the linear propagation approach proposed by
Fiorentini \emph{et al.~}\cite{Fiorentini1998} and then
generalized by Cuoco \emph{et al.~}\cite{Cuoco2004} to estimate
the error matrix. The observational abundances and uncertainties
of $^{4}\text{He}$, $\text{D}$ and $^{7}\text{Li}$ are taken from
Luridiana \emph{et al.~}\cite{Luridiana2003}, Kirkman \emph{et
al.~}\cite{Kirkman2003} and Bonifacio \emph{et
al.~}\cite{Bonifacio2002} respectively as:
\begin{eqnarray}
\text{Y}_{\text{P}}^{\text{obs}} &=& 0.2391 \pm 0.0020;\\
(\text{D/H})^{\text{obs}} &=& 2.78_{-0.38}^{+0.44} \times
10^{-5};\\
(^{7}\text{Li}/\text{H})^{\text{obs}} &=&
2.19_{-0.38}^{+0.46}\times 10^{-10}
\end{eqnarray}
(Note that the $^4\text{He}$ abundance is quantified by its mass
fraction.) These observational results are also shown in Figs.~1
and 2 as grey regions, and in Fig.~1 we also show the earlier
wider range of $^{4}\text{He}$ given by Olive, Steigman and Walker
\cite{Olive2000b},
\begin{equation}
\text{Y}_{\text{P}}^{\text{obs}} = 0.238 \pm 0.005,
\end{equation}
by the dark grey region as a comparison, even though we will not
use it in our likelihood analysis.

The results for $r=10$ are plotted in the upper panel of Fig.~3,
where we treat $\eta$ and $\rho$ as two free parameters and give
the 68\% C.L.~contours from the abundances of $\text{D}$,
$^{4}\text{He}$ and $^{7}\text{Li}$ individually. It is apparent
that with the presence of an evolving radion field (or
equivalently a varying Higgs VEV so that $\rho\neq1$) the
observational abundances of these 3 elements can be compatible at
the $1\sigma$ level and consistent with the WMAP result. This
qualitative conclusion is then confirmed by the lower panel of
Fig.~3, in which the joint constraints from
$\text{D}$+$^{4}\text{He}$+$^{7}\text{Li}$ at 68\% and 95\% C.L.~
are shown. Our best-fitting point in the parameter space (the
white circle) is at $\rho\simeq1.009$ and
$\eta\simeq6.28\times10^{-10}$. This central value of $\eta$ is
slightly larger than that of $\eta_{\text{WMAP}}$
($\eta_{\text{WMAP}}=6.14^{+0.25}_{-0.25}\times10^{-10}$) but
falls into the $1\sigma$ range of the latter. In contrast, the SBBN
(horizontal line) is only allowed marginally at 95\% C.L.~ and
should be further excluded at the same level with
$\eta_{\text{WMAP}}$ taken into account. A similar situation is
presented for $r=6$ (Fig.~4), while the details are a little
different: here the best-fitting point $\sim(6.67\times10^{-10},
1.02)$ lies slightly outside the $1\sigma$ range of
$\eta_{\text{WMAP}}$ and SBBN is excluded at the 95\% C.L.~ using
the observational primordial abundances only. In any case we see
an improvement compared with SBBN if $\rho$ is appropriately
chosen (note also that, if the $1/B_{d}$ scaling of the cross sections
discussed in Sec.~II B is a good approximation, then $\rho>1$ will also
increases the rates of these reactions by reducing $B_{d}$, making
the destructions of $\text{D}$ more efficient; this will lead to
less $\text{D}$ yields and shift the $\text{D}$ and joint contours
leftward).

In some cases we may be interested in how $\rho$ alone is allowed
to change, and for this purpose it proves to be convenient to use
the constraint on $\eta$ from the WMAP measurement (the
$\eta_{\text{WMAP}}$ given above) by virtue of its high accuracy.
We shall construct a likelihood function of $\rho$ only as
\cite{Dmitriev2004}:
\begin{equation}
L(\rho)\propto\int^{\infty}_{-\infty}\text{exp}\left[-\frac{(\eta-\eta_{0})^2}
{2\sigma^{2}_{\eta}}\right]\text{exp}\left[-\frac{\chi^2(\eta,\rho)}
{2}\right]d\eta,
\end{equation}
in which $\eta_{0}$ and $\sigma_{\eta}$ are respectively the
central value and $1\sigma$ range for $\eta$ given by WMAP;
$\chi(\eta,\rho)$ is a function of both $\eta$ and $\rho$
calculated using the same method as above (for
$\text{D}$+$^4\text{He}$+$^7\text{Li}$). The results are shown in
Fig.~5, where we give the cases for both $r=10$ and $r=6$ and we
have found that the 95\% C.~L.~ranges for $\rho$ in these two
cases are $\rho=1.0089\pm0.0012$ and $\rho=1.0182\pm0.0024$,
corresponding respectively to changes in the extra space volume
$V$ as:
\begin{equation}
1.52\times10^{-2}\lesssim\frac{\Delta
V}{V_0}\lesssim1.99\times10^{-2}
\end{equation}
and
\begin{equation}
3.09\times10^{-2}\lesssim\frac{\Delta
V}{V_0}\lesssim4.00\times10^{-2},
\end{equation}
where $\Delta V=V_{0}-V_{\text{BBN}}$ is the difference between
the extra space volumes at present and at the BBN era. The allowed
variation of $\rho$ (or equally the Higgs VEV) we find here is
comparable to the value quoted in \cite{Yoo2003}, but the full
likelihood analysis using the $\eta_{\text{WMAP}}$ constraint and
the new observational abundances of $\text{D}$, $^4\text{He}$ and
$^7\text{Li}$ presented here strongly disfavor $\rho \leq
1$. Furthermore, as seen from Fig.~5, the ranges for $\rho$ with
$r=10$ and $r=6$ do not agree with each other at the 95\% C.~L.,
even though in both cases the qualitative features are the same.
This simply reflects the demand of a more precise understanding
of how the deuteron binding energy depends on the light
quark masses.

If no constraint on $\eta$ is used, we obtain looser constraints
on $\rho$, namely $\rho=1.0091\pm0.0071$ for $r=10$ and
$\rho=1.0198\pm0.0140$ for $r=6$ (again at the 95\% C.L.), which
are comparable to the results above. In this case a variation of
the Higgs VEV as large as $\sim3.4\%$ compared with its present
value is still allowed at the BBN era (redshift
$z\sim10^9-10^{10}$).

\section{DISCUSSIONS AND CONCLUSIONS}

In summary, we have considered in this article the possible
implications of a cosmologically evolving Brane moduli field on
the outputs of the primordial nucleosynthesis. We begin with the
discussions on how the fundamental constants are modified if the
radion field of the Brane varies and how these modifications would
affect the BBN yields. Some effects not considered previously are
included in this procedure. Then we present a likelihood analysis
using the recent compilation of the various nuclear reaction rates
\cite{NACRE1999} and observational abundances of the light nuclei
$\text{D}$, $^4\text{He}$ and $^7\text{Li}$ \cite{Luridiana2003,
Kirkman2003, Bonifacio2002}. The WMAP constraint on $\eta$
\cite{Spergel2003} is also adopted in determining the constraint
on the quantity $\rho$, which characterizes the variation of the
Higgs VEV (and thus the evolution of the radion field in the
present scenario). We find that the BBN yields could be changed in
such a way that the discrepancy between SBBN and
$\eta_{\text{WMAP}}$ might be reduced, provided that the Higgs VEV
was slightly larger at the BBN era. This conclusion is robust
within the present theoretical uncertainty of $r$ ($6\sim10$), and the errors
introduced by neglecting the $\rho$-dependences of other nuclear reactions
are estimated to well fall within the $1\sigma$ $\text{D}$, $^4\text{He}$
and $^7\text{Li}$ observational uncertainties. However,
further developments in the understanding of these topics
might help constrain $\rho$
more accurately. The constraints we obtain in this work are also
applicable to other models in which the Higgs VEV (and fundamental
fermion masses) varies cosmologically, and could be used to
constrain parameters in such models.

It would be also interesting to consider other implications of
such an evolving radion field. One example is its influences on
the CMB; this was discussed in Refs.~\cite{Yoo2003, Kujat2000} and
the authors found that these influences were mainly through the
variation of the electron mass $m_{e}$ (see \cite{Kujat2000} for
more details). This might provide another constraint on the radion
field, but it is much looser than the one we obtain and at a
different cosmic era ($z\sim10^3$). Another potential implication
of interest is the possible existence of di-proton or di-neutron.
This is because the bindings of the di-nucleon systems are mainly
contributed by the long range pion force which depends on pion
mass $m_{\pi}$. If $m_{\pi}$ was smaller at the time of BBN, then
the bindings of di-proton/di-neutron would be enhanced, and if the
di-proton becomes bound, it will open a rapid channel for the
hydrogen fusion \cite{Dyson1971} and thus be catastrophic to the
stellar lifetimes. The conditions for the di-proton and di-neutron
binding were studied in \cite{Dent2003, Barrow1987}. In the
present work, we see that $\rho>1$ is preferred by our BBN
analysis, which means that $m_{\pi,\text{BBN}}>m_{\pi,\text{NOW}}$
and the di-nucleon systems would be kept unbound during the BBN
era. As the last example of possible implications, let us consider
the stability of $^5\text{He}$ which, if becoming stable, would
fill the mass-5 gap in the BBN nuclear reaction chain, drastically
enhancing the production of $^7\text{Li}$. In \cite{Flambaum2002}
the authors investigated this question and found that
$\delta(m_{q}/\Lambda_{\text{QCD}})
/(m_{q}/\Lambda_{\text{QCD}})\gtrsim-0.1$ was required to prevent
$^5\text{He}$ from becoming stable, which is obviously satisfied
by our constrained results.

\begin{acknowledgments}
The work described in this paper was partially supported by a
grant from the Research Grants Council of the Hong Kong Special
Administrative Region, China (Project No.~400803).
\end{acknowledgments}

\appendix

\newcommand{\noopsort}[1]{} \newcommand{\printfirst}[2]{#1}
  \newcommand{\singleletter}[1]{#1} \newcommand{\switchargs}[2]{#2#1}


\begin{thebibliography}{19}
\bibitem[{\citenamefont{Comment2}(2005)}]{Comment2}
  One may consider that there are $1/B$ dependences of the forward
cross sections like the $1/B_{d}$ dependence of the $p(n,
\text{D})\gamma$ cross section discussed above, $B$ being the relevant binding
energies. If this is the case,
then there will be another error for the $^7\text{Li}$ yield of
order a few percents, but it is hard to estimate it more exactly as
we do not know explicitly how these cross sections depend on $\rho$.
\end{thebibliography}
\end{document}